\documentclass{iaus}
\usepackage{graphicx}

\title[Multiple Population Theory: Extreme helium population problem] 
{Multiple Population Theory:\\ Extreme helium population problem}

\author[Sukyoung K. Yi]   
{Sukyoung K. Yi$^1$}

\affiliation{$^1$ Yonsei University, Department of Astronomy, Seoul 120-749, Korea \break email: yi@yonsei.ac.kr}

\pubyear{2008}
\volume{258}  
\pagerange{??--??}
\date{?? and in revised form ??}
\jname{Proceedings Title IAU Symposium}
\editors{A.C. Editor, B.D. Editor \& C.E. Editor, eds.}
\begin{document}

\maketitle

\begin{abstract}
The spreads in chemical abundances inferred by recent precision observations
suggest that some or possibly all globular clusters can no longer be
considered as simple stellar populations. The most striking case is
$\omega$~Cen in the sense that its bluest main-sequence despite its high
metallicity demands an extreme helium abundance of $Y \approx 0.4$.
I focus on this issue of ``the extreme helium population problem'' 
in this review.
\end{abstract}

\firstsection 

\section{Introduction}

Globular clusters may not be a robust example for $simple$ stellar populations
any more. Perhaps there is no such thing as simple stellar populations
from the beginning. 
The classic globular clusters, such as $\omega$~Cen, NGC\,2808, and
NGC\,1851 are now suspected to be composed of heterogeneous populations,
and recent data from the space with unprecedented resolving accuracy
are hinting at a great fraction of Milky Way globular clusters 
being composite populations at least chemically.

At the centre of debates is $\omega$~Cen.
It has long been known as a mysterious object.
To begin with, the spectacular southern cluster is the most massive in 
Milky Way, with some million solar mass.
Its unusually broad red giant branch was found to indicate {\em discrete}
multiple populations by the magnificent effort and insight of 
Lee et al. (1999) using the mere 0.9m telescope.
More recent works with much superior instruments unambiguously revealed the 
multiplicity of the giant cluster.
Norris (2004) and Bedin et al. (2005) sequentially found that the 
multiplicity is evident not just in the red giant branch but also in the 
main sequence.
To everyone's surprise, its bluest main sequence is too blue for the
metallicity of $\omega$~Cen and in fact more metal rich
than the redder main-sequence stars (Piotto et al. 2005).
If such a blue colour for such a metal-rich population is real,
it unavoidably indicates possibility of the scorchingly high
helium abundance, $Y \approx 0.4$. 
The blue main-sequence population constitutes 30\% in number (Bedin et al. 
2004; Sollima et al. 2007) and thus not something we can simply sweep under 
the carpet.
If there is any good news in this apparent nightmare
the blue main-sequence population seems to be at least younger than the
majority of the stars in this cluster, perhaps by a couple of billion years 
(Villanova et al. 2007). 

Significant is its implication to the extended horizontal branch in this
cluster. This and many other clusters exhibit an extended horizontal branch,
and its origin has been a long-debated issue.
Apparently, the same level of extreme helium inferred by the blue main
sequence explains the extreme blueness of the extended horizontal branch
as well (Lee et al. 2005).
If this prevails in other clusters as well, the hitherto mysterious 
origin for the extended horizontal branch may also be solved by the extreme
helium.

Apparently many more clusters show multiple sequences, either on the main
sequence and/or on the sub-giant branch (Piotto 2008, this volume), even
though it is not yet clear whether such multiplicities are also to be
interpreted as originated from extreme helium.
More massive clusters tend to 
show multiple sequences more often, and interestingly
the same trend is found for the extended horizontal branch (Recio-Blanco et al.
2006; Lee et al. 2007).

Extragalactic counterparts to $\omega$~Cen and the like may have been
found in the giant elliptical galaxy M87 in the Virgo cluster
as well (Sohn et al. 2006; Kaviraj et al. 2007).
Using the Hubble Space Telescope Sohn et al. (2006) found that most of
the massive globular clusters in M87 are UV bright despite their likely 
old ages, as if they have an extremely hot horizontal branch.
Through an extensive test using the UV-focused population synthesis models 
of Yi (2003), Kaviraj et al. (2007) concluded that the UV
strengths (a tracer of the horizontal branch morphology) of these clusters 
are even stronger than that of $\omega$~Cen by more than a magnitude.
Whatever is causing the mutiplicity to $\omega$~Cen seems to affect
the M87 clusters even more greatly.

Massive clusters showing various anomalies seems to corroborate the
idea of them originally being something of a different nature, for example, nucleated dwarf spheroidal galaxies (Lee et al. 1999).

All things considered, there appears to be a huge conspiracy.
It is not yet clear whether the multiplicity in the main sequence is
the cause of that in the horizontal branch. But, they all fit in a very
sensible storyline.
Although it ruins the old and naive concept of ``simple stellar populations'', 
multiplicity itself is perhaps not a huge problem. 
The extreme helium abundance inferred by the blue main sequence population
would be an exciting discovery to observers but a desperate-to-forget nightmare to theorists. 
I discuss why that is.

\section{Significance}

The significance of this issue is immense.
First, understanding how such an extreme helium abundance is possible is
an interesting challenge. 
It also influences the current age dating techniques that are based on
the precise main sequence fitting and on detailed horizontal branch
analysis. 
The seemingly-settled issue on the second parameter problem of horizontal
branch may enter a new stage with the not-so-new idea of helium with 
a clearer understanding on helium enrichment processes.
The endless debate on the origin of the UV upturn found in bright
elliptical galaxies may find a new and compelling explanation with helium.
Obvious too is the impact on the issue of the age of the universe, as
globular clusters and bright elliptical galaxies are often considered
the oldest stellar populations in the universe. 

\section{Observational facts and inferences}

Finding a solution to the case of $\omega$~Cen is only a beginning step
because other clusters show different constraints, but it would still 
be a good start. So I attempt to find a solution adopting some of the most
widely-discussed channels. 

Our {\em simplified} constraints are as follows.
\begin{itemize}
\item The age separation: the blue main sequence subpopulation is 
1--3 billion years younger than the red main sequence subpopulation; 
i.e., $t(bMS) \approx t(rMS) - 1$--3 (Lee et al. 2005; Stanford et al. 2007; 
Villanova 2007). I think the exact value is poorly constrained but 
for now take $\Delta t =1$Gyr as a face value.
\item The mass fraction: the number (and mass) fraction of the blue main 
sequence subpopulation is roughly 30\% (Bedin et al. 2004), 
i.e., $f(bMS) = 0.3$. I will try to aim to find a solution to satisfy this. 
However, this may not place as strong a constraint as I take it, 
if the mass evolution of sub-populations is complex. 
I will discuss this in detail in \S 6.
\item Discrete sub-populations: the main sequence and horizontal branch 
splits appear very sharp and discrete. Hence, a stochastic element 
in a solution to the extreme helium abundance cannot be dominant. 
Instead, it has to offer a process that leads to a clear prediction in 
helium abundance.
\item The metal abundance: $Z(rMS)=0.001$ and $Z(bMS)=0.002$ 
(Piotto et al. 2005). The metallicity of the blue main sequence is 
difficult to pin down due to their faintness and so still uncertain. 
But it seems clear that it is higher than that of the red main sequence.
\item The helium abundance: the helium abundance of the blue main sequence 
sub-population is 40\% in mass, i.e., $Y=0.4$. In reality, the observed 
colour-magnitude diagramme shows even up to 5 sub-populations. 
But it is impossible to make a model that pins down all the sub-populations 
found. Hence, I approximate them into 2 sub-populations: the red main 
sequence has an ordinary helium abundance and the blue main sequence has
an extreme helium abundance. As I will discuss in the end, it is 
perhaps very important to remind ourselves repeatedly that {\em the 
helium abundance was never directly measured but inferred from the 
main sequence fitting.} Despite this, I take it as a face value.
\item The helium enrichment parameter: the helium and metal abundances 
lead to the incredible helium enrichment parameter, 
$\Delta$$Y$/$\Delta$$Z \approx 70$. Ordinary populations with ordinary stars 
yield $\Delta$$Y$/$\Delta$$Z \approx 2$--3 even for a wide variance of 
initial mass functions. Hence, this poses the most challenging problem of 
all. I will focus pretty much of my tests on this issue.
\item Other elements: spectroscopic measurements on carbon and nitrogen 
are available, i.e., $[C/M]\sim 0$ and $[N/M] \sim 1$ for the 
blue main sequence population. However, their accuracy seems not as 
good as one might hope for and the error estimations (i.e. measurement 
significance) are not provided. It is already a daunting task to reproduce 
the helium properties, and so I will only use this information as a reference.
\end{itemize}

\section{Asymptotic giant branch stars}

The most obvious candidate origin for such an extreme helium is asymptotic
giant branch stars (e.g., Izzard et al. 2004; D'Antona et al. 2005 among many). 
Although there is quite a scatter in the chemical yield prediction, 
there is consensus that asymptotic giants generate a copious amount of
helium but only a small amount of metals (e.g., Maeder 1992). 
This is good to us because we do not just want to produce a lot of helium 
but want to achieve the high value of helium enrichment parameter as well.
Supernovae for comparison produce too much metals to satisfy this
constraint, although they are also good producers of helium. 
This is such a basic understanding that it does not require elaboration, but 
it has recently been pointed out in quantitative matter anyway 
(Choi \& Yi 2007).

The asymptotic giants in a narrow mass range ($M \approx 5$--6) indeed
release ejecta of the high value of helium enrichment parameter that we aim 
to achieve. So if a population receives the stellar mass ejecta mainly
from asymptotic giant stars but nothing else, it is in principle possible
to achieve such a high value of helium enrichment parameter. 
More massive stars would produce both metals and helium.
Hence, an {\em ad hoc} scenario, where all the mass ejecta from massive stars
(say $M>M_{\rm esc}$ where $M_{\rm esc} \sim 5$--10 solar mass) would 
escape the gravitational potential where a subsequence star formation
occurs, can be set up to maximise the impact of the asymptotic giants
in terms of the helium enrichment parameter.
The effectiveness of the {\em maximum AGB scenario} has been discussed
by a few groups (e.g., Karakas et al. 2006; Bekki et al. 2007), and
Choi \& Yi (2008) performed a detailed calculation to check its viability.

Choi \& Yi (2008)  adopted a toy model where the original gas reservoir
does not accept any new gas infall from outside but the mass ejecta
from massive stars above $M_{\rm esc}$ escape it supposedly via
supernova explosion, hence maximising the helium enrichment effect 
from asymptotic giants. It is plausible that the kinetic energy
of the mass ejecta from supernova explosion achieve the
escape energy of such a small potential well.
It is assumed that a fraction (50 -- 100\%) of the initial gas is 
used to form the first (red main sequence) population and the
subsequent population (blue main sequence) will be born from the 
remnant gas mixed with the stellar mass ejecta from the first population.
The abundance of the initial gas is assumed to be the abundance of the
red main sequence population of $\omega$~Cen.
If a higher fraction of the initial gas reservoir is used to build
the first population, it would obviously result in a higher value of
helium abundance and helium enrichment parameter for the second
population, but only a small amount of gas becomes available for
the second population formation.  

If we do not adopt any constraint on the age difference between the
red and blue main sequence populations, we can achieve a very high
helium abundance ($Y\approx 0.36$ which is almost as high as we aime
to reach) and the maximum value of helium enrichment parameter of 
about 70 as we hoped for.
In this case, the age difference is roughly 0.1Gyr, and the second
generation is virtually a pure recycling product of the first generation
stellar mass ejecta within a narrow mass range of 5--6 solar mass.
But in this case, the total mass ratio between the red and blue main
sequence populations becomes 99.3: 0.7; that is, only 0.7\% of the
total population in $\omega$~Cen can benefit from this scenario.
Since the blue main sequence population is observed to be 30\% instead
0.7\%, there is a factor of 40 discrepancy! I call this ``the mass
deficit problem''.

One may achieve somewhat different estimates by adopting different
yields. For example, Renzini (2008) uses the recent yield
for the so-called ``super-AGB stars'' to find that the mass discrepancy
can be as small as 15 instead of 40. 

If we take the age difference of roughly 1Gyr as a valid constraint,
the situation becomes dramatically worse.
This is because, even if we assume the $M_{\rm esc}$ argument,
the stars in the mass range 2--5 solar mass will now contribute 
to the gas reservoir through stellar mass ejecta which is in general
of substantially lower helium abundance ($\sim 0.3$) 
and helium enrichment parameter ($\sim 2$--5).
Consequently, this scenario with 1Gyr age separation can achieve up to
$Y \approx 0.3$ and $\Delta$$Y$/$\Delta$$Z \approx 10$ while the
upper limit in the mass fraction of the second generation is just 7\%
(instead of 30\%).
Let alone the shortcoming in the helium properties, the mass
fraction requirement cannot be met, either.

The verdict on the maximum AGB scenario and its variation can be summarised
as follows. The extreme helium-related propertis are almost impossible if 
the age difference is a meaningful constraint, hence making this
scenario totally implausible.
If the age separation constraint can be eased off, the extreme
value of the helium enrichment parameter ({\em but not the helium abundance
itself}) can be reproduced by the first generation of 
asymptotic giants for the following conditions and criticisms.
\begin{itemize}
\item The stellar mass ejecta from massive stars of $M > 6$ must all escape
the gravitational potential well. If the super AGB scenario (e.g. Siess
2007) is adopted, this mass limit can be as high as 10 solar mass.
If all supernova ejecta leave the system as high wind material, this is
not a bad assumption, but assuming that the supernova ejecta leave
completely without affecting the remaining gas in the reservoir
is extreme and very unlikely. 
\item The blue main sequence population must form exactly after 0.1Gyr
after the red main sequence population, in disagreement with the
1Gyr separation suggested by previous studies. I personally think
the age separation constraint is not strong and thus
0.1Gyr is not particularly unappealing. 
\item The mass decifit of a factor of 40 (which can be somewhat smaller
if super AGB stars are adopted) is a serious threat and requires
a rescue plan.
A possible remedy may be found in the details of the cluster dynamical
evolution, which is discussed in \S 6.
\item An encouraging aspect of this scenario is that the discreteness of the
separated populations is easy to explain. The second generation forms
from the mass loss of the first generation 0.1Gyr later. 
\end{itemize}

\section{Fast-rotating massive stars}

A totally different solution was put forward by the massive stellar evolution
models. Maeder \& Meynet (2006) suggested that metal-poor 
massive stars that are rotating
nearly at their break-up speed may release a lot of helium via 
{\em slow wind}
before they start burning heavy elements and explode as a supernova.
Their idea came from their earlier works (Maeder \& Meynet 2001)
that suggested (1) low metallicity stars reach break-up rotational speed
more easily by the combined effect of stellar (slow) winds and rotation,
(2) they have efficient mixing of their core materials, that is, 
helium and other heavier elements (depending on the rotation speed)
out to the surface, and (3) during their blue
loop after the red giant phase, a fast contraction leads to excessive
mass loss from the helium and nitrogent enhanced surface material.

The elemental yields via slow wind are sensitive to the rotational speed 
adopted. For example for a 60 solar mass star with $log Z = -5$,
a fast rotating model at 85\% of the break-up speed yields
the helium abundance of 5.86 solar mass, the metal abundance of
0.09 solar mass, and thus the helium enrichment parameter of
63.3 (which is very close to our aim!). On the other hand,
a moderately fast rotating model at 35\% of the break-up speed,
the yields become $\Delta Y = 1.73$, $\Delta Z = 2.6e-5$, and
$\Delta$$Y$/$\Delta$$Z \approx 10^5$. These extremely fast-rotating
stars generate excessively high values of $\Delta$$Y$/$\Delta$$Z$ and
too little of helium. 
The fast rotating stars overproduce carbon and nitrogen abundances 
compared to observation, while the moderate
rotating stars reproduce the observation better. But we still select the
fast-rotating models in our exercise (Choi \& Yi 2007) 
because they produce much more helium and thus more likely to satisfy our aim.

The toy model of Choi \& Yi (2007) using the metal-poor massive rotating
stars of Maeder \& Meynet (2006) show that a simple population based
on an ordinary initial mass function indeed achieves
the high values of helium abundance and of $\Delta$$Y$/$\Delta$$Z$ 
in the stellar mass loss, as we aim to recover. These values are further 
elivated by the helium-dominant contribution from asymptotic
giants until lower mass giants become the main source of chemical yields.
Thus this phenomenon of high helium properties lasts only for a 
short period of time of order 0.1--0.2Gyr, just as in the AGB scenario.
Once the population becomes older than that, its stellar mass loss
accumulated will no longer have such high values of helium properties.

We find, however, that the mount of the gas with the high helium properties 
can be only roughly 1.4\% of the total stellar mass of $\omega$~Cen
which is a factor of 20 too small for it to be the sole solution to
this problem.
This mass deficit of a factor of 20 is smaller (and thus better) than 
that of the asymptotic giant branch star scenario simply because this time
we have helium contributions from massive rotating stars as well as
from asymptotic giants.
Here, we assume that only the slow wind stellar mass loss from the massive
stars remain in the gravitating system and the fast wind (explosions)
materials leave the system without polutting the gas reservoir.

In conclusion, we could not find a solution if the age separation of
1 Gyr or so is a meaningful constraint. For a much smaller age
separation of order 0.1Gyr, we could achieve high values of helium
parameters, but even in this case the mass available for the formation
of the second generation is a factor of 20 smaller than what we have
in $\omega$~Cen. This problem has been noted also by a much more
detailed dynamical simulation of Decressin, Baumgardt, \& Kroupa (2008;
see also the article by Decressin in this volume).
We will discuss this further in \S 6.

Another serious problem in this scenario is the carbon abundance.
While it depends strongly on the rotational speed adopted,
the 60 solar mass model with 85\% of the break-up speed suggests
that the slow wind mass loss will be highly enriched in carbon, which
is not supported by the observational data (Piotto et al. 2005).

For this scenario to be appealing, we also need to understand how 
a specific rotation speed is determined for the stars. Why does it happen
to some clusters (like $\omega$~Cen) but not to others?
Is it randomed given to each cluster, and not to each forming star?
That will be very odd. 
This scenario with fast-rotating mass stars certainly adds to what
was already possible from the asymptotic giant stars and thus provides
a positive contribution. However, it alone does not appear to provide 
a full solution to our problem.

\section{Dynamical evolution}

The blue main sequence population seems more centrally concentrated
than the red main sequence population. If this was true from the start
one can expect that the spatially more extended red main sequence 
population will lose more stars throughout its dynamical evolution.
D'Ercole et al. (2008; and also the poster at this meeting) 
indeed suggested that a substantial fraction of the first generation of stars 
may escape the system if some conditions are met. 
For example, if the initial mass distribution within each globular 
cluster follows the King profile and {\em if its King radius is equal to
its true tidal radius}, then it is very easy to shed some high velocity stars
into space.
In this case, only 2-3\% of the original first generation stars
may remain in the cluster mainly due to the kinetic energy injection by 
supernova explosions and two-body relaxation.
If this is true, it makes both the AGB scenario and the massive
fast-rotating star scenario viable.
 
Whether these conditions were easy to meet by the first generation 
clusters is not yet clear, however. More traditional studies (e.g., Fall \&
Zhang 2001) based on evaporation by two-body relaxation, gravitational shocks,
and stellar mass loss suggest an order of magnitude milder mass evolution.
 
The mass evaporation is supposed to be sensitive to the mass of the 
cluster in the sense that {\em a more massive cluster would shed less mass}.
So, if the dynamical evaporation was indeed the key to this extreme helium
phenomenon, it would be very unlikely to happen preferrentially to the most
massive clusters.
Unfortunately to this scenario, $\omega$~Cen is the Milky Way's most
massive cluster and the other clusters showing multiplicity, NGC\,2808
and M\,54, are among the most massive, too.
Besides,  the extended horizontal branch globular clusters in the Milky Way
and the UV-brightest clusters in M\,87 all occupy the highest
mass end in the total cluster mass distribution of the galaxy.
In this sense, the dynamical evaporation picture loses its charm.

If D'Ercole et al's dramatic mass evolution is applicable to
all globular clusters, then it would have a significant impact on the 
cluster luminosity function evolution. Typical clusters in the Milky Way 
are of a million solar mass presently which is in the same order as the 
size of the giant molecular
clouds, the main site of star formation, and as the mass of the
star clusters forming in nearby mergeing galaxies.
In this regard, I feel that this scenario of shedding 98\% of the 
initial mass of the first population is likely a rare event. 
Perhaps this is why the main sequence splits are not a common feature.
Otherwise, that is, if such a dramatic mass evaporation had been
true to all clusters, then we should find our galactic stellar halo 
to have at least ten billion solar masses, which is an order or magnitude
greater than the current estimate.
I strongly feel that phyiscal understanding on the dynamical 
process (when such conditions are met) is required, and detailed 
dynamical models, cross-checking with
the observed cluster lumnosity functions, are called for.

\section{The first stars}

There must have been stars before population I and II stars.
This is evident from several arguments.
Theoretically, the mass of the first objects that experience dynamical
instability is estimated to be stellar rather than galactic.
This is consistent with the fact that reionisation is (although indirectly) 
observed through the cosmic microwave background radiation studies.
Observationally, despite the fact that the big bang itself did not 
generate any appreciable amount of heavy elements, 
totally metal-deficient stars are not found anywhere.
Even the most metal-deficient stars show $log Z \sim -5$ and more
typical metal-poor stars have metallicity greater than a hundredth
of the solar value.
This means the pregalactic gas reservoir must have been enriched in metals
substantially.
The most probable objects for this are the first stars, a.k.a. population
III stars.
The first stars are often thought to have been very massive, above
a hundred solar mass, while other possibilities are also being considered.

The duration of the first star formation episode is considered to be
extended well into that of population II (Bromm \& Loab 2006).
If we are considering a proto-galactic scale system, the mixing time
scale for the chemical elements may have been of order hundred million
years, and thus considering both the extended star formation timescale
and varying mixing timescale, some {\em chemical inhomegeneity} in 
the gas reservoir for the population II star formation was inevitable. 

Marigo et al. (2003) have computed the chemical yields for such metal-free
stars of mass between 100 and 1000 solar mass.
Surprisingly their models suggest that the first stars were very efficient
in generating and releasing helium into the space but not metals.
This is mainly because the first stars had such an enormous radiation pressure
that the balance between the mass accretion for growing up and 
radiative pressure was difficult to achieve; that is, 
the strong radiation pressure blew away the gas that was being accreted.
So the first energy generation involving hydrogen burning was possible
but before the star reaches the next stage it would release much
materials processed by then: i.e., helium.
This results in a high helium to metal ratio, as we were looking for.

Choi \& Yi (2007) have indeed investigated this effect to the 
helium enrichment in the gas cloud.
They found that the the range between 100 and 1000 solar masses,
a lower-mass first star produces a much larger value of 
$\Delta$$Y$/$\Delta$$Z \sim 10^{7-8}$. (No, this is not a typo.)
First stars of 1000 solar mass are predicted by this model to have
$\Delta$$Y$/$\Delta$$Z \sim 10^{2}$, which is much closer to our aim.
Adopting a Salpeter initial mass function\footnote{As I type this part
I just learned of Professor Salpeter's death. We have just lost one of
the greatest astrophysicists of our time.}, 
we found that a first star 
population with a mass range 100---1000 solar mass releases virtually 
no metals but abundant helium, and thus reaching
$\Delta$$Y$/$\Delta$$Z \sim 500$. 
A population with a higher value of the lower bound results in a 
gradually lower value.
Eventually, a population purely made up of 1000 solar-mass stars would
have $\Delta$$Y$/$\Delta$$Z \approx 70$. 

After letting the first star population evolve for a billion years
the remnant gas cloud (primordial gas left out of the first star formation
plus the stellar mass loss mixed evenly) reaches the metallicity
of the blue main sequence ($Z=0.002$), the hulium abundance ($Y=0.4$), and
so the helium enrichment parameter ($\Delta$$Y$/$\Delta$$Z \approx 70$), 
with no further free parameter.

This scenario is briefly investigated by Choi \& Yi (2008) and 
can be chronologically described as follows.
\begin{enumerate}
\item The majority of first stars form in the universe at redshift roughly 
at 20 ($t \equiv 0$).
\item These stars release much helium and some metals.
\item The chemical mixing in the proto-galactic cloud took a long time, 
and after hundreds of million years, chemically-mixed regions are more 
common than unmixed regions.
\item From a chemically mixed region, the red main-sequence population of 
$\omega$~Cen forms ($t \sim 0.5$Gyr).
\item From the pristine (unmixed) gas in the vicinity a second generation
of first stars form ($t \sim 0.7$Gyr).
\item They generate abundant helium and little metals and enrich the remnant
gas cloud to $Z \sim 0.002$, $Y \sim 0.4$ and thus 
$\Delta$$Y$/$\Delta$$Z \sim 70$. 
\item From this gas cloud, the blue main-sequence population of $\omega$~Cen
forms ($t \sim 1.5$Gyr).
\item The blue main-sequence population merges into the more massive
red main-sequence population soon after their formation.
\end{enumerate} 

This picture is very rough however and contains many caveats.
\begin{itemize}
\item The first star chemical yields may be highly uncertain.
A more robust understanding on the formation and evolution on first stars
will perhaps come in the near future, but more importantly independent
calculations (besides Marigo et al.) would be required immediately.
\item In this scenario the first stars (at least the ones that led to the
gas reservoir for the formation of the blue main-sequence population)
should have very high mass of order 1000 solar mass.
This is not supported by some recent first star studies.
\item We need not just a couple of first stars in this region but more than
one hundred. How such a material gathers up in this proto cloud is
a mestery, especially when first stars are often believed to form
isolated rather than in multiplets.
\item The physics in terms of the chemical mixing and its timescale is
highly uncentain, as is the case for other scenarios.
\end{itemize}
Given all these uncertainties, it is difficult to argue that the first
scenario is any more compelling than others. However, it is still 
a very exciting possibility. After all, we astronomers are always
the first one to find something wrong as well as new and important.
This conjecture at least implores for more studies on the first stars.

\section{Alternative theories}

Alternative theories are also available.
The velocity-dispersion dependent surface pollution of AGB ejecta scenario
was put forward by Tsujimoto et al. (2007).
A similar surface pollution scenario was presented by Newsham \& Turndrup
(2007). While the channels for the pollution can be several,
it provides an interesting possibility that the blue main-sequence stars
are not truly so helium-rich as we believe but pretend to be so
by having unusually high helium abundance only on the stellar surface.
Mass transfer of the surface material in binary stellar systems could
be one channel, or if stars passing through the cluster central region
where helium-rich gas from the accumulated stellar mass loss is located
such stars may be polluted on the surface.
However, it is very unlikely that 30\% of the stars get contaminated
like this. Besides, all these processes would occur in random manner that
the discreteness of the blue main-sequence would be unnatural.

A possibility of having primordial fluctuations in the helium density 
was presented by Chuzhoy (2006). In this study, the helium diffusion time
scale for the primordial gas of stellar mass was of order of hundred 
million years, and thus some of the birth clouds for first stars were
heavily enriched in helium. But again, the diffusion timescale
must depend on each birthcloud condition which should be rather random,
which makes the main sequence discreteness a tough problem to solve.

\section{Conclusions}

Theoriests are often very optimistic thinking that a tough problem
to solve is challenging instead of mind-boggling. But I must admit
that I am much more than puzzled by this ``extreme helium population problem''.
The presence of multiple populations in globular cluster size populations
is surely a problem, as numerical simulations of the kind performed
by Bate, Bonnell, \& Bromm (2008) suggest that the star formation
in a cluster probably happens in the crossing time scale,
which is an order of million years.
But we have seen other small populations having a complex star formation
history, e.g., Carina dwarf galaxy (Smecker-Hane et al. 1994).
A more critical issue is the extreme value of helium properties.
I do not believe that we have a compelling theory yet.
Asymptotic giant branch stars are a familiar class and thus makes
our mind susceptible. But I believe that I have shown that it still
has the mass deficit problem by a factor of at least 40, which is
threateningly large even to astronomers.
Same is true for the massive stars rotating nearly at the break-up 
speed. They alone cannot provide the full answer and suffer from 
a similar mass deficit problem.
Its physical plausibility is also something to be worked out.
The first star scenario is fascinating because first stars are
a mystery in general. We believe that they were once around 
but have never seen them, a bit like black holes.
They provide a plausible solution, but just barely.
It has so many caveats and uncertainties that cannot be clarified 
in the next few years. Hence it loses its charm, too.

I said at the end of my presentation at this conference that 
the enigmatic extrme helium population is so tough to theorists
that I would almost feel happy if someone comes up to say 
``It was all a mistake from the start. There is no such extreme
helium population''. George Meynet disagreed with me. He instead
said the problem is so enigmatic that we are greatly challenged and
excited. I became humble at his constructive attitudes. I hope to
see a more believable solution in the near future.

\begin{acknowledgments}
I thank Ena Choi for countless constructive discussions. 
Much of this review is based on several key papers written by Choi \& Yi 
(2007, 2008), Decressin, Charbonnel \& Meynet (2008), and by Renzini (2008). 
I am grateful to Young-Wook Lee, Suk-Jin Yoon, Ken Nomoto (during my visit
to Tokyo University), and Enrico Vesperini for constructive discussions.
Special thanks are due to Changbom Park for stimated discussion on 
stellar collisions in clusters during my visit to the KIAS.
This research has been supported by Korea Research Foundation (SKY).
\end{acknowledgments}


\begin{thebibliography}{}


\bibitem[\protect\citeauthoryear{Bedin et al.}{2005}]{b04}
Bedin, L. R., Piotto, G., Anderson, J., Cassisi, S., King, I. R., Momany, Y.,  \& Carraro, G. 2004, ApJ, 605, L125
\bibitem[\protect\citeauthoryear{Bekki et al.}{2007}]{b07}
Bekki, K., Campbell, S. W., Lattanzio, J. C. \& Norris, J. E., 2007,
MNRAS, 337, 335
\bibitem[\protect\citeauthoryear{Choi \& Yi}{2007}]{cy07}
Choi, E. \& Yi, S. K. 2007, MNRAS, 375, L1
\bibitem[\protect\citeauthoryear{Choi \& Yi}{2008}]{cy08}
Choi, E. \& Yi, S. K. 2008, MNRAS, 386, 1332
\bibitem[\protect\citeauthoryear{Chuzhoy}{2006}]{chu06}
Chuzhoy, L. 2006, MNRAS, 369, L52
\bibitem[\protect\citeauthoryear{D'Antona et al.}{2005}]{dan05}
D'Antona, F., Bellazzini, M., Caloi, V., Pecci, F.  F.,  Galleti, S.,  \& Rood, R. T. 2005, ApJ, 631, 868
\bibitem[Decressin et al. 2007]{dec07} Decressin, T., Charbonnel, C., \& Meynet, G. 2007, A\&A, 475, 859
\bibitem[Decressin et al. 2008]{dec08} Decressin, T., Baumgardt, H., \& Kroupa, P. 2008, A\&A, in press
\bibitem[Fall \& Zhang 2001]{fz01} Fall, S. M. \& Zhang, Q. 2001, ApJ, 561, 751
\bibitem[\protect\citeauthoryear{Izzard et al.}{2004}]{Izz04}
Izzard, R. G., Tout, C. A., Karakas, A., \& Pols, O. R. 2004, MNRAS, 350, 407
\bibitem[\protect\citeauthoryear{Karakas et al.}{2006}]{ka06}
Karakas, A. I., Fenner, Y., Sills, A., Campbell, S. W., \& Lattanzio, J. C. 2006, ApJ, 652, 1240
\bibitem[\protect\citeauthoryear{Kaviraj et al.}{2007}]{kav07}
Kaviraj, S., Sohn, S. T., O'Connell, R. W., Yoon, S.-J, Lee, Y.-W., \& Yi, S. K. 2007, MNRAS, 377, 987
\bibitem[\protect\citeauthoryear{Lee et al.}{1999}]{lee99}
Lee, Y.-W., Joo, J.-M., Sohn, Y.-J., Rey, S.-C., Lee, H.-C., \& Walker, A. R. 1999, Nature, 402, 55
\bibitem[\protect\citeauthoryear{Lee et al.}{2005}]{lee05}
Lee, Y.-W., Joo, S.-J., Han, S.-I., Chung, C., Ree, C. H., Sohn, Y.-J., Kim, Y.-C., Yoon, S.-J., Yi, S. K. \& Demarque, P. 2005, ApJ, 621, L57
\bibitem[\protect\citeauthoryear{Lee et al.}{2007}]{lee07}
Lee, Y.-W., Gim, H. B., \& Casetti-Dinescu, D. I. 2007, ApJ, 661, L49
\bibitem[\protect\citeauthoryear{Maeder}{1992}]{m92}
Maeder, A. 1992, A\&A, 264, 105
\bibitem[\protect\citeauthoryear{Maeder \& Meynet}{2001}]{mm01}
Maeder, A. \& Meynet G. 2001, A\&A, 373, 555
\bibitem[\protect\citeauthoryear{Maeder \& Meynet}{2006}]{mm06}
Maeder, A. \& Meynet G. 2006, A\&A, 448, L37
\bibitem[\protect\citeauthoryear{Newsham \& Terndrop}{2007}]{nt07}
Newsham, G. \& Terndrop, D.M. 2007, ApJ, 664, 332
\bibitem[\protect\citeauthoryear{Norris}{2004}]{n04}
Norris, J. E. 2004, ApJ, 612, 25
\bibitem[\protect\citeauthoryear{Piotto et al.}{2005}]{pi05}
Piotto et al. 2005, ApJ, 621, 777
\bibitem[\protect\citeauthoryear{Piotto et al.}{2007}]{pi07}
Piotto, G., Bedin, L. R., Anderson, J., King, I. R., Cassisi, S., Milone, A. P., Villanova, S., Pietrinferni, A. \& Renzini, A. 2007, ApJ, 661, 53
\bibitem[\protect\citeauthoryear{Recio-Blanco~et~al.}{2006}]{rb06}
Recio-Blanco, A., Aparicio, A., Piotto, G., de Angeli, F., \& Djorgovski, S. G.
2006, A\&A, 452, 875
\bibitem[Renzini 2008]{ren08} Renzini, A. 2008, MNRAS, 391, 354
\bibitem[Siess 2007]{s07} Siess, L. 2007, A\&A, 476, 893
\bibitem[Smecker-Hane et al. 1994]{s94} Smecker-Hane, T.A., Stetson, P.B., Hesser, J.E., \& Lehnert, M.D. 1994, AJ, 108, 507
\bibitem[\protect\citeauthoryear{Sohn et al.}{2006}]{sohn06}
Sohn, S. T., O'Connell, R. W., Jundu, A., Landsman, W.B., Burstein, D.,
Bohlin, R.C., Frogel, J.A., \& Rose, J.A. 2006, AJ, 131, 866
\bibitem[\protect\citeauthoryear{Sollima et al.}{2007}]{sol07}
Sollima, A., Ferraro, F. R., Ballazzini, M., Origlia, L., Straniero, O., \&
Pancino, E. 2007, ApJ, 654, 915
\bibitem[\protect\citeauthoryear{Stanford et al.}{2007}]{sta07}
Stanford, L. M., Da Costa, G. S., Norris, J. E., \& Cannon, R. D.
2007, ApJ, 667, 911
\bibitem[\protect\citeauthoryear{Tsujimoto, Shigeyama, \& Suda}{2007}]{Tsu07}
Tsujimoto, T., Shigeyama, T., \& Suda, T. 2007, ApJ, 654, 139
\bibitem[\protect\citeauthoryear{Villanova et al.}{2007}]{vil07}
Villanova, S., Piotto, G., King, I. R., Anderson, J., Bedin, L. R., Gratton, R. G., Cassisi, S., Momany, Y., Bellini, A., Cool, A. M.
Recio-Blanco, A., \& Renzini, A. 2007, ApJ 663, 296
\bibitem[\protect\citeauthoryear{Yi}{2003}]{yi03}
Yi, S. 2003, ApJ, 582, 202

\end{thebibliography}
\end{document}